\documentclass[aps,prb,twocolumn,superscriptaddress,showpacs,
preprintnumbers,amssymb, 10pt]{revtex4-2}

\usepackage{graphicx}
\usepackage{color}
\usepackage[normalem]{ulem}

\newcommand{\Mn}{Li$_2$Sr[MnN]$_2$}
\newcommand{\Mnx}{Li$_2$Sr[(Li$_{1-x}$Mn$_x$)N]$_2$}

\newcommand{\eps}{\varepsilon}
\newcommand{\sg}{$I\,4_1/a\,m\,d$}

\begin{document}
\title{{Li$_2$Sr[MnN]$_2$}: a magnetically ordered, metallic nitride}

\author{F. Hirschberger}
\affiliation{EP VI, Center for Electronic Correlations and Magnetism, Institute of Physics, University of Augsburg, D-86135 Augsburg, Germany}
\author{T. J. Ball\'e}
\affiliation{EP VI, Center for Electronic Correlations and Magnetism, Institute of Physics, University of Augsburg, D-86135 Augsburg, Germany}
\author{C. Haas}
\author{W. Scherer}
\affiliation{CPM, Institute of Physics, University of Augsburg, D-86135 Augsburg, Germany}
\author{A. A. Tsirlin}
\affiliation{EP VI, Center for Electronic Correlations and Magnetism, Institute of Physics, University of Augsburg, D-86135 Augsburg, Germany}
\author{Yu. Prots}
\author{P. H\"ohn}
\affiliation{Max-Planck-Institut f\"ur Chemische Physik fester Stoffe, N\"othnitzer Str. 40, D-01187 Dresden, Germany}
\author{A. Jesche}
\affiliation{EP VI, Center for Electronic Correlations and Magnetism, Institute of Physics, University of Augsburg, D-86135 Augsburg, Germany}

\date{\today}
\begin{abstract}
\Mn~single crystals were successfully grown out of Li rich flux. 
The crystal structure was determined by single crystal X-ray diffraction and revealed almost linear $-$N$-$Mn$-$N$-$Mn$-$ chains as central structural motif.
Tetragonal columns of this air and moisture sensitive nitridomanganate were employed for  electrical transport, heat capacity, and anisotropic magnetization measurements. 
Both the electronic and magnetic properties are most remarkable, in particular the linear increase of the magnetic susceptibility with temperature that is reminiscent of underdoped cuprate and Fe-based superconductors. 
Clear indications for antiferromagnetic ordering at $T_N = 290$\,K were obtained.
Metallic transport behavior is experimentally observed in accordance with electronic band structure calculations.
\end{abstract}


\maketitle

\section{Introduction}
Oxides are normally associated with ionic bonding and insulating electrical transport behavior.
When this intuition fails, particularly interesting physical properties and ground states can be found.
Well known examples are ferrites above the Verwey transition\,\cite{Walz_2002}, cuprate superconductors\,\cite{Timusk1999}, or delafossites\,\cite{Putzke2020}.
In general, nitrides tend to behave similar to oxides, however, metallic behavior is arguably even more rare.  

Binary transition metal nitrides attract considerable interest due to their valuable mechanical, electrical and magnetic properties. 
In contrast, chemistry and physics of multinary nitrides have been far less thoroughly explored\,\cite{Hohn2017}.
Nitridometalates of $d$ metals $T$ represent an interesting class of solid state phases, which contain nitrogen as isolated anions N$^{3-}$ or feature complex anions [T$_x$N$_y$]$^{z-}$ of different dimensionalities with coordination numbers of $T$ by N typically between two and four and oxidation states of the transition metals being comparatively low. 
Whereas the bonding within these complex anions and frameworks is essentially covalent, nitridometalates are stabilized by predominantly ionic bonding through counterions of electropositive metals like alkali ($A$) or alkaline-earth ($AE$) cations.
\\

Nitridomanganates show a wide variety of structures and properties: manganese may be coordinated tetrahedrally (Li$_7$[Mn$^\mathrm V$N$_4$]\,\cite{Niewa2001}) or trigonal-planar 
(Li$_{24}$[Mn$^\mathrm{III}$N$_3$]$_3$N$_2$\,\cite{Niewa2001}, 
Ca$_6$[Mn$^\mathrm{III}$N$_3$]N$_2$\,\cite{Gregory1995}, 
Ca$_3$[Mn$^\mathrm{III}$N$_3$]\,\cite{Tennstedt1993}, 
Sr$_3$[Mn$^\mathrm{III}$N$_3$]\,\cite{Tennstedt1993b, Barker1996}, 
Ba$_3$[Mn$^\mathrm{III}$N$_3$]\,\cite{Tennstedt1993b,Barker1996}, 
Sr$_8$[Mn$^\mathrm{III,IV}$N$_3$]$_3$\,\cite{Bendyna2008}) in isolated units; ethane-like units are observed in Li$_6$Ca$_2$[Mn$^\mathrm{IV}_2$N$_6$]\,\cite{Hochrein1998} and Li$_6$Sr$_2$[Mn$^\mathrm{IV}_2$N$_6$]\,\cite{Hochrein2003}. 
Charge ordering is observed in chains of edge-sharing tetrahedra in Ba$_4$[Mn$^\mathrm{II,IV}_3$N$_6$]\,\cite{Ovchinnikov2018ba4} whereas complex 2D structures of mixed valent manganese in both tetrahedral and trigonal-planar coordination are observed in Ca$_{12}$[Mn$^\mathrm{II,III}_{19}$N$_{23}$] and Ca$_{133}$[Mn$^\mathrm{II,III}_{216}$N$_{260}$]\,\cite{Ovchinnikov2018ca12}.
Both of those semiconducting phases order antiferromagnetically; their localized magnetism originates from only a small fraction of the Mn atoms not involved in  metal-metal bonding and is characterized by a large orbital contribution. 
Anti-rutile-type (Mn$_{2-x}$Li$_x$)N\,\cite{Niewa1998} features 3D networks of trigonal-planar coordinated manganese. 
Linear substituted chains [(Li$_{1-x}$Mn$_x$)N] are observed in Li$_2$[(Li$_{1-x}$Mn$^\mathrm I_x$)N]\,\cite{Niewa2001,Klatyk1999d}, Li$_5$[(Li$_{1-x}$Mn$^\mathrm{I,II}_x$)N]$_3$\,\cite{Niewa2001} and Li$_2$Ca[(Li$_{1-x}$Mn$_x$)N]$_2$\,\cite{Klatyk2000}.

The single crystal growth of nitrides is often challenging due to the large dissociation energy of the N$_2$ molecule, the reactivity of the starting materials and enhanced vapor pressures, but the high temperature centrifugation aided filtration (HTCAF) technique\,\cite{Bostrom2001,Jesche2014c} using Li as a flux provides a powerful tool to produce large single crystals of a large variety of phases even in complex systems such as nitridometalates (e. g. Li$_2$Sr[(Li$_{1-x}$Fe$^\mathrm I_x$)N]$_2$\,\cite{Hohn2016}, LiSr$_2$[Co$^\mathrm I$N$_2$]\,\cite{Balle2019}) or nitride metalides like LiSr$_3$Ga$_2$N\,\cite{Pathak2017}.
Although magnetic properties were reported for only few nitridometalates containing alkaline-earth metals, these new phases show high magnetic anisotropy (Li$_2$Sr[(Li$_{1-x}$Fe$^\mathrm I_x$)N]$_2$\,\cite{Hohn2016}) and ferromagnetic (LiSr$_2$[Co$^\mathrm I$N$_2$]\,\cite{Balle2019}) or antiferromagnetic (LiSr$_2$[Fe$^\mathrm I$N$_2$]\,\cite{Hoehn2021}) ordering.
\\

%
%
%

Here we present a metallic nitride based on the ionic alkali-alkaline-earth-nitrogen host Li$_4$SrN$_2$ substituted with Mn. 
Since Mn almost fully occupies one of the two Li sites of Li$_4$SrN$_2$ and is absent on the other one, we refer to the material as \Mn.
The phase is isotypic to Li$_2$Sr[(Li$_{1-x}$Fe$^\mathrm I_x$)N]$_2$, but shows a significantly different magnetic behavior\,\cite{Hohn2016}.  
Single crystals of several millimeter along a side can be obtained from Li-rich flux. 
We find clear indications for magnetic ordering close to room temperature and metallic electrical transport behavior.


\begin{figure*}
\includegraphics[width=0.96\textwidth]{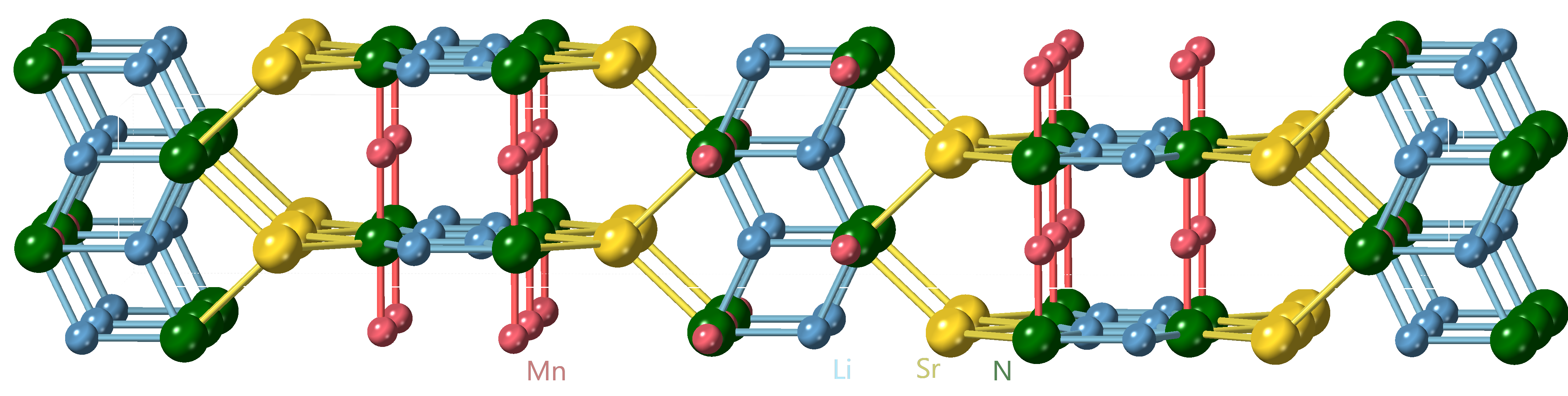}
\caption{\label{CrystStruc} 
Crystal structure of \Mn, the unit cell is indicated by the grey lines (space group \sg). Mn is substituted in linear, two-fold coordination between N. One set of the $-$N$-$Mn$-$N$-$Mn$-$ chains is oriented vertical, the other one roughly perpendicular to the paper plane.
}
\end{figure*}

\section{Methods}

Starting materials were lithium rod (Evochem, 99\,\%), manganese powder (Alfa Aesar, 99.9998\,\%), strontium nitride (Sr$_2$N) powder [prepared from strontium metal (Alfa Aesar, distilled dendritic pieces, 99.95\,\%) and nitrogen (Westfalengas, 99.999\,\%, additionally purified by molecular sieves and BTS catalyst)] and sodium azide (NaN$_3$) powder (Roth, 99\,\%) as a further nitrogen source. 
Tantalum ampules were produced on-site from pre-cleaned tantalum tube and tantalum sheet (Plansee) in an arc furnace located within a glove box.

Laue back-reflection pattern were taken with a digital Dual FDI NTX camera manufactured by Photonic Science (tungsten anode, $U = 20$\,kV). The calculated pattern as well as the picture of the crystal structure were created using CrystalMaker and SingleCrystal.
Laboratory powder X-ray diffraction data of finely ground (gray) powder samples were collected on a Huber G670 imaging plate Guinier camera using a curved germanium (111) monochromator and Cu-K$\alpha$1 radiation in the range $4^\circ \leq 2\theta \leq 100^\circ$ with an increment of $0.005^\circ$ at 293(1)\,K. 
The powder samples were placed between Kapton foils to avoid degradation in air.
All preparations and sample handling were performed under inert atmosphere (Ar).
Phase identification was done using the WinXPow program package\cite{XPow2003}.

The crystal structures and the compositions of several crystals obtained in different experiments of the title compound \Mnx~were refined from single-crystal X-ray diffraction data.
Data for single crystals 1,2, and 4 were collected at temperatures 400 K, 300 K, and 100 K, respectively, on a Bruker APEX2 diffractometer with a D8 goniometer and Ag-K$\alpha$ ($\lambda = 0.56087$\,\AA) employing Helios mirror optics (Incoatec). 
The crystal was kept at 100 K and 400 K using a Cryostream 700 low temperature device (Oxford Cryosystems) during data collection. 
Using Olex2\,\cite{Dolomanov2009} the structure was solved with the ShelXS\,\cite{Sheldrick2008} structure solution program using Direct Methods and refined with the ShelXL\,\cite{Sheldrick2015} refinement package using Least Squares minimisation. 

Data for single crystal 3 was collected at room temperature on a Rigaku AFC7 four circle diffractometer equipped with a red{Saturn 724+} CCD detector (Mo-K$\alpha$ radiation, graphite monochromator). 
For numerical absorption correction, an optimized shape of the single crystal was employed\,\cite{stoe1999data,stoe1999shape}.
After data collection the structure was confirmed by direct methods, using SHELXS\,\cite{Sheldrick2008} and subsequently refined by using the full-matrix least-squares procedure with SHELXL\,\cite{Sheldrick2015}.
Further details on the crystal structure investigations may be obtained from the Fachinformationszentrum Karlsruhe, 76344 Eggenstein-Leopoldshafen, Germany (Fax: +49-7247-808-666; e-mail: crysdata@fiz-karlsruhe.de), on quoting the depository numbers CSD-xxxxxx, the names of the authors, and the journal citation.

The magnetization was measured using a 7\,T Magnetic Property Measurement System (MPMS3) manufactured by Quantum Design. 
The electrical transport properties were measured as a four-probe measurement with spring loaded contacts using a 14\,T Physical Properties Measurement Systems which was also manufactured by Quantum Design.
The side of the few millimeter sized samples on which the contacts were placed was ground off under argon atmosphere before each run.
This served to remove insulating layers from the surface of the crystals, which are sensitive to air and moisture.
Due to the limited size of the available crystals and the measurement setup, it was not possible to investigate the anisotropy of electrical transport.
The data shown in this work were obtained from AC electrical transport measurements at a frequency of 117\,Hz and 9\,Hz.
The specific electric resistivity was calculated from the measured resistance based on the sample geometry by following the procedure described in\,\textcite{Eder1972}. 

Density-functional (DFT) band-structure calculations were performed in the \texttt{FPLO} code~\cite{Koepernik1999} using local density approximation (LDA) for the exchange-correlation potential~\cite{Perdew1992}. Correlation effects in the Mn $3d$ shell were introduced on the mean-field DFT+$U$ level with the double-counting correction in the atomic limit. The on-site Hund's coupling was fixed to $J_H=1$\,eV, while the Coulomb repulsion parameter $U$ was varied between 1 and 10\,eV to explore the influence of correlation effects on the electronic structure. Magnetic moments in spin-polarized calculations are estimated by the internal routine of \texttt{FPLO} via projections on local atomic orbitals. Experimental structural parameters at 300\,K (data set 2b) were used in all calculations.


\begin{figure}
\includegraphics[width=0.44\textwidth]{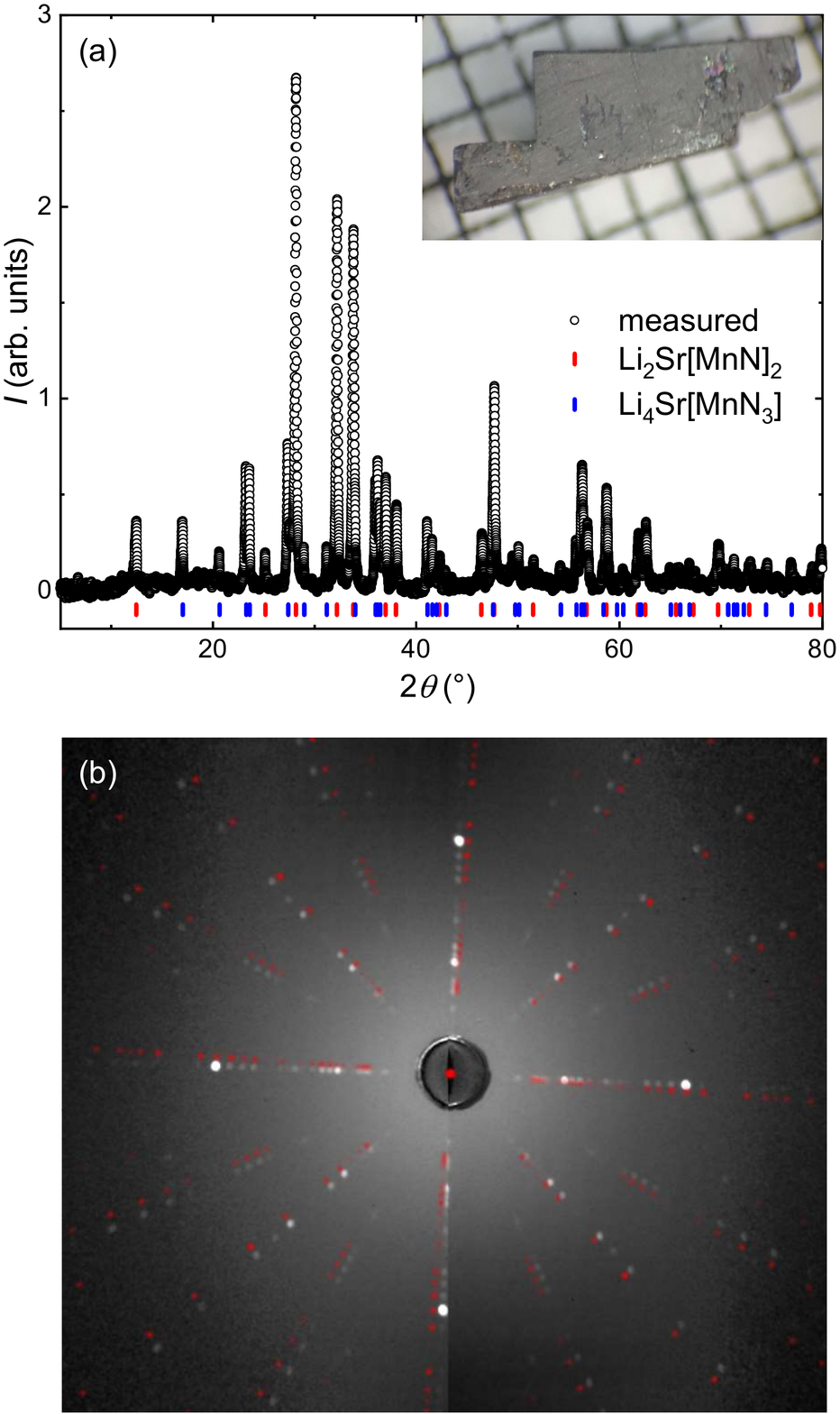}
\caption{\label{XRay} X-ray diffraction pattern measured after grinding part of the product obtained by Li flux growth. Both \Mn~and Li$_4$Sr[MnN$_3$] were identified (Cu-K$\mathrm\alpha$ radiation, $\lambda = 1.54056\,$Å). The inset shows a single crystal of \Mn~on a millimeter grid. b) Laue back-reflection pattern recorded along the $c$-axis of the \Mn~single crystal shown above. The calculated pattern (red spots) is slightly rotated for clarity}.
\end{figure}

\section{Single crystal growth}

Due to the air and moisture sensitivity of both the reactants (Li, NaN$_3$, Sr$_2$N) and the final product Li$_2$Sr[(Li$_{1-x}$Mn$_x$)N]$_2$ all manipulations including grinding and weighing, as well as complete sample preparations for measurements were carried out in an inert gas glove box (Ar, O$_2$ and H$_2$O$\leq 1$\,ppm).
Large single crystals of \Mnx~in the form of square columns with a silverish luster were obtained by reaction of Sr$_2$N, Mn, NaN$_3$, and Li in molar ratio 10\,:\,34\,:\,1\,:\,340 with NaN$_3$ acting as a nitrogen source and Li as flux and mineralizer. 

The mixtures with total mass of about 0.6\,g were placed into tantalum ampules equipped with a sieve\,\cite{Canfield2001}, sealed by arc welding under inert atmosphere of 700 mbar argon and subsequently encapsulated in a quartz tube with an internal argon pressure of 300 mbar in order to prevent oxidation of the tantalum. 
The samples were heated from room temperature to $T = 1023$\,K with a rate of 100\,Kh$^{-1}$, annealed for 2\,h, subsequently cooled to 600\,K with 1\,Kh$^{-1}$ and further with 2\,Kh$^{-1}$ to 400\,K. 
After centrifugation with 3000\,min$^{-1}$ employing the HTCAF method\,\cite{Bostrom2001,Jesche2014c} to separate the single crystals from the excess flux. 

Large single crystals of the title compound in form of tetragonal columns with a silverish black lustre were obtained in addition to greenish black single crystals of the new phase Li$_4$Sr[MnN$_3$] and a small amount of hexagonal reddish black Li$_3$N as side phases. 
Using a two-probe multimeter, the differently colored phases showed remarkably different resistivities further enabling the discrimination of the various phases. 
Small single crystals suitable for X-ray diffraction were obtained either directly selected from the product or by crushing larger specimens.
One of the largest \Mn~single crystals obtained is shown as an inset of Fig.\,\ref{XRay}(a).

The HTCAF technique using a Li-rich flux has been successfully employed for the growth of large single crystals of Mn-substituted Li$_4$SrN$_2$. 
However, it was not possible to obtain single phase material using this reaction route up to now. Although no binary intermetallic phases seem to be stable under these conditions, other nitrides and nitridometalates may form in this quinary (Li-Sr-Mn-N-Ta) system depending on composition, temperature and reaction time. Whereas the ratio AE/N significantly influences product formation, the amount of Li as flux seems to play only a minor role. 
During short-term reactions, excess N results in formation of highly oxidized transition metals or Li$_3$N, whereas after long reaction times also reactions with the crucible materials come into play as evidenced by the formation of Li$_7$[TaN$_4$]. 
Our attempts to synthesize strontium nitridomanganates yielded a large variety of phases such as Li$_2$Sr[(Li$_{1-x}$Mn$^\mathrm I_x$)N]$_2$, Li$_4$Sr[Mn$^\mathrm{III}$N$_3$], or Sr$_3$[Mn$^\mathrm{III}$N$_3$]. 
Whether large single crystals of other known nitridomanganates as well as phases with Mn in other oxidation states can be grown by adjusting temperature profile and/or the ratio of the starting materials is subject of ongoing research. 
Altogether, the HTCAF method may prove a useful tool for a wide range of applications. 

\begin{table*}
\caption{\label{TabCrystData} Crystallographic data and experimental details for the single crystal structure refinements of different crystals of Li$_2$Sr[(Li$_{1-x}$Mn$_x$)N]$_2$.}
\begin{tabular}{|c|c|c|c|c|}\hline 
sample&1&2&3&4\\\hline
\bf{\textit{T} [K]}		&\bf{400}	&\bf{300}	&\bf{295}(2)	&\bf{100}\\\hline
\bf{Mn occupancy (x)}	&\bf{0.954(5)}	&\bf{0.946(3)}	&\bf{0.931(5)}	&\bf{0.949(4)}\\\hline
crystal system								&tetragonal&tetragonal&tetragonal&tetragonal\\\hline
space group									&\sg~(No. 141)&\sg~(No. 141)&\sg~(No. 141)&\sg~(No. 141)\\\hline
$a$ [\AA]		&3.8151(3)	&3.8115(4)	&3.8126(2)	&3.8101(6)\\\hline
$c$ [\AA]		&28.445(2)	&28.347(3)	&28.372(3)	&28.245(5)\\\hline
$V$ [\AA$^3$]	&414.02(7)	&411.82(10)	&412.41(6)	&410.03(14)\\\hline
$Z$				&4			&4			&4			&4\\\hline

$\rho_\mathrm{calc}$ [gcm$^{-3}$]
				&3.769		&3.778		&3.740		&3.800\\\hline
crystal color, habit							&metallic, black, block&metallic, black, block&silverish, tetragonal column&metallic, black, block\\\hline
crystal size [mm]							&0.175$\times$0.153x0.115&0.223$\times$0.16$\times$0.107&0.15$\times$0.18$\times$0.20&0.192$\times$0.185$\times$0.126\\\hline
$\mu$ [mm$^{-1}$]							&9.791&9.821&18.264&9.875\\\hline
sin\,$\theta_{\rm max}/\lambda$ [1/$\lambda$]		&1.00&1.00&0.89&1.00\\\hline
diffractometer								&BRUKER APEX2&BRUKER APEX2&RIGAKU AFC7&BRUKER APEX2\\\hline
wavelength [Å]								&0.56087 (Ag-K$\alpha$)&0.56087 (Ag-K$\alpha$)&0.71069 (Mo-K$\alpha$)&0.56087 (Ag-K$\alpha$)\\\hline
monochromator								&mirror optics&mirror optics&graphite&mirror optics\\\hline
scan mode									&$\omega$ scans&$\omega$ scans&$\varphi$ scans&$\omega$ scans\\\hline
measured reflections							&13690&7790&3266&8813\\\hline
observed reflections							&525&509&370&512\\
$[I_\mathrm{o} > 2\sigma(I_\mathrm{o})]$		&&&&\\\hline
$R_\mathrm{int}$								&0.0517&0.0418&0.0765&0.0508\\\hline
number of parameters							&16&16&16&16\\\hline
goodness-of-fit on $F^2$						&1.345&1.216&1.161&1.274\\\hline
wR2											&0.0616&0.0554&0.0981&0.0659\\\hline
R1 [$F_\mathrm{o} > 4\sigma(F_\mathrm{o})$]	&0.0320&0.0249&0.0376	&0.0300\\\hline
R1 (all data)								&0.0354&0.0291&0.0507&0.0332\\\hline
residual electron density 					&1.31/-1.57&0.920 /-0.788&1.66/-2.01&1.520/-1.343\\
$[e\times 10^{-6}\,\mathrm{pm}^{-3}]$&&&&\\\hline

\end{tabular} 
\end{table*}

\section{Crystal structure}

\begin{table}
\caption{\label{TabAtomicPos} Atomic positions and equivalent displacement parameters [Å$^2$] for Li$_2$Sr[(Li$_{1-x}$Mn$_x$)N]$_2$ for crystals 1-4. The Li occupancy on the {\bf Mn}Li site is obtained by $1-x$(Mn), all other parameters for this site were constrained to equal values for Mn and Li.}
\begin{tabular}{|c|c|c|c|c|c|}\hline 
&$x$&$y$&$z$&$U_\mathrm{eq}$&$x(\mathrm{Mn})$\\\hline

Sr1&1/2&1/4&1/8&0.01016(9)&\\
&&&& 0.00838(7)&\\
&&&& 0.01286(19)&\\
&&&&0.00420(8)&\\
\hline

{\bf Mn}Li&0&3/4&0.05213(2)&0.01076(13)&0.954(5)\\
              &&&0.05210(2)&0.00844(10)&0.946(3)\\
              &&&0.05209(3)&0.0118(3)  &0.931(5)\\
              &&&0.05213(2)&0.00470(11)&0.949(4)\\\hline


N1&1/2&3/4&0.05444(10)&0.0091(4)&\\
        &&&0.05452(8) &0.0076(3)&\\
        &&&0.05455(15)&0.0128(7)&\\
        &&&0.05450(9) &0.0051(3)&\\\hline

Li1&1/2&3/4&-0.0225(3)&0.0156(13)&\\
         &&&-0.0226(2)&0.0128(9)&\\
         &&&-0.0224(4)&0.0182(18)&\\
         &&&-0.0223(3)&0.0096(11)&\\\hline
 
\end{tabular} 
\end{table}

\begin{table}
\caption{\label{TabAniDisp} Anisotropic displacement parameters [Å$^2$] for Li$_2$Sr[(Li$_{1-x}$Mn$_x$)N]$_2$ for crystals 1-4. ($U_{12} = U_{13} = U_{23} = 0$, Li and Mn on the {\bf Mn}Li site were treated identical).}
\begin{tabular}{|c|c|c|c|}\hline 
Atom&$U_{11}$&$U_{22}$&$U_{33}$\\\hline
Sr1&0.01062(12)&&0.00923(15)\\
   &0.00836(9)&&0.00843(12)\\
   &0.0114(2)&$U_{11}$&0.0157(3)\\
   &0.00404(10)&&0.00451(13)\\\hline

{\bf Mn}Li&0.00363(18)&0.0145(2)&0.0142(2)\\
          &0.00300(13)&0.01057(16)&0.01174(18)\\
          &0.0057(3)&0.0131(4)&0.0165(5)\\
          &0.00195(17)&0.00571(19)&0.00643(19)\\\hline


N1&0.0039(8)&0.0100(10)&0.0134(9)\\
  &0.0039(6)&0.0078(7)&0.0110(7)\\
  &0.0109(16)&0.0111(16)&0.0165(18)\\
  &0.0032(8)&0.0047(9)&0.0076(8)\\\hline

Li1&0.021(4)&0.012(3)&0.014(3)\\
   &0.016(2)&0.009(2)&0.013(2)\\
   &0.015(4)&0.018(4)&0.022(4)\\
   &0.012(3)&0.006(3)&0.010(2)\\\hline

\end{tabular} 
\end{table}
\begin{table}
\caption{\label{TabDistAngl} Relevant distances (in \AA) and angles in Li$_2$Sr[(Li$_{1-x}$Mn$_x$)N]$_2$ [\AA] for crystals 1-4. }
\begin{tabular}{|c|c|c|c|c|c|}\hline 
&&1&2&3&4\\\hline
Sr1-N1&4x&2.769(2)&2.761(2)&2.762(3)&2.756(2)\\\hline
N1-MnLi&2x&1.9087(2)&1.9070(2)&1.9076(2)&1.9062(3)\\\hline
N1-Li1&2x&2.113(4)&2.110(3)&2.113(5)&2.111(3)\\\hline
N1-Li1&1x&2.189(8)&2.185(6)&2.184(11)&2.170(7)\\\hline
$\angle$N1-MnLi-N1&&176.0(2)&175.9(2)&175.8(3)&176.0(2)\\\hline

\end{tabular} 
\end{table}

Figure~\ref{XRay} shows the X-ray powder diffraction pattern measured on a ground mixture of single crystals of Li$_2$Sr[(Li$_{1-x}$Mn$_x$)N]$_2$ and Li$_4$Sr[MnN$_3$]. 
Lattice parameters of $a = 3.8087(3)$\,\AA~and $c = 28.334(4)$\,\AA~obtained from the Li$_2$Sr[(Li$_{1-x}$Mn$_x$)N]$_2$ powder differ somewhat from single crystal data (see below) indicating a variation of the Mn concentration within the samples. 
The decrease in $a$ by 0.34\,\% and increase in $c$ by 4.8\,\% with increasing Mn content $x$ in comparison to the parent compound Li$_4$SrN$_2$ ($a = 3.822(2)$\,\AA, $c = 27.042(9)$\,\AA\,\cite{Cordier1989b}) is in line with the trend reported for the isotypic iron 
($x = 0.32$, $a = 3.79536(9)$\,\AA, $c = 27.6492(13)$\,\AA\,\cite{Hohn2016}; 
$x = 0.46$, $a = 3.7909(2)$\,\AA, $c = 27.719(3)$\,\AA\,\cite{Klatyk1999b}), 
cobalt ($x = 0.3$, $a = 3.7363(3)$\,\AA, $c = 27.9063(24)$\,\AA\,\cite{Hoehn2021}), 
nickel ($x = 0.1$, $a = 3.823(1)$\,\AA, $c = 27.074(5)$\,\AA\,\cite{Cordier1989b}), 
and copper phases ($x = 0.39$, $a =  3.770(1)$\,\AA, $c = 27.386(6)$\,\AA\,\cite{Jager1992}). 

Smaller crystals were obtained from several different crushed samples and selected for single crystal X-ray diffraction at several temperatures. 
The results are summarized in Tables\,\ref{TabCrystData}\,-\,\ref{TabDistAngl}.
Li$_2$Sr[(Li$_{1-x}$Mn$_x$)N]$_2$ crystallizes tetragonal in space group \sg~and is a substitution variant of Li$_4$SrN$_2$\,\cite{Cordier1989b}.


Predominant structural feature of Li$_2$Sr[(Li$_{1-x}$Mn$_x$)N]$_2$ are 2D fragments of the Li$_3$N structure with composition Li[(Li$_{1-x}$Mn$_x$)N]. 
These slabs are connected by tetrahedrally coordinated Sr atoms and stacked  along [001] in such a way that each layer is rotated 90\,$^\circ$ to the previous one. 
Powder as well as single crystal X-ray diffraction revealed slightly varying Mn-concentrations of $x = 0.94\pm 0.02$; the slight differences between powder and single crystal data may also stem from slight inhomogeneities within the samples.
As indicated by the notation of the chemical formula, Li$_2$Sr[(Li$_{1-x}$Mn$_x$)N]$_2$, the substituted Mn atoms occupy only one of the two Li-sites: 
whereas there is no evidence for substitution on the trigonal planar coordinated Li position, the almost linearly ($\angle$N1-MnLi-N1 = 176(0.2)$^\circ$, see Table\,\ref{TabDistAngl}) coordinated chain position [(Li$_{1-x}$Mn$_x$)N] is nearly completely substituted by Mn ($x = 0.94(2)$, see Table\,\ref{TabAtomicPos}).
Compared to a truly linear chain, the (Li$_{1-x}$Mn$_x$) atoms are deflected towards the Li[(Li$_{1-x}$Mn$_x$)N] slabs. 
A designation as Li$_2$Sr[MnN]$_2$ is also conceivable and used in this article, since on average only every 18th manganese atom within the chains is substituted by lithium.
Laue back-reflection pattern calculated based on the obtained structure are in excellent agreement with measurements recorded on large single crystals [Fig.\,\ref{XRay}(b)].

The structural feature of linear substituted chains [(Li$_{1-x}$Mn$_x$)N] is also observed in 
Li$_2$[(Li$_{1-x}$Mn$^\mathrm I_x$)N]\,\cite{Niewa2001,Gregory1995}, 
Li$_5$[(Li$_{1-x}$Mn$^\mathrm{I,II}_x$)N]$_3$\,\cite{Niewa2001} and 
Li$_2$Ca[(Li$_{1-x}$Mn$_x$)N]$_2$\,\cite{Klatyk2000}.
Whereas distances $d$(Li$_{1-x}T_x$-N) decrease significantly with increasing $x$ for $T$ = Fe, Co, Ni\,\cite{Kniep2013x} and to a certain extent Cu\,\cite{Jager1992}, this effect is less pronounced in nitridomangantes: 
$d$(Li$_{1-x}$Mn$_x$-N) = 1.9076(2)\,\AA~in Li$_2$Sr[(Li$_{0.07}$Mn$_{0.92}$)N]$_2$ ,
1.911\,\AA~in Li$_5$[(Li$_{0.33}$Mn$^\mathrm{I,II}_{0.67}$)N]$_3$\,\cite{Niewa2001}, 
1.914\,\AA~in Li$_2$[(Li$_{0.27}$Mn$^\mathrm I_{0.73}$)N]\,\cite{Klatyk1999d} and 
1.916\,\AA~in Li$_2$Ca[(Li$_{0.06}$Mn$_{0.94}$)N]$_2$\,\cite{Klatyk2000} compared to $d$(Li-N) = 1.938\,\AA~in Li$_3$N\,\cite{Rabenau1976} and 1.913\,\AA~in Li$_4$SrN$_2$\,\cite{Cordier1989b}.

Considering all isotypic substitution variants, the manganese phase is a special case, as there is almost complete substitution ($x = 0.94(2)$) of the linearly coordinated lithium position (compared to obtained substitution levels in
Li$_2$Ca[(Li$_{1-x}$Fe$_x$)N]$_2$: $x = 0.3$\,\cite{Klatyk1999b},
Li$_2$Sr[(Li$_{1-x}$Fe$_x$)N]$_2$: $x = 0.46$\,\cite{Klatyk1999b},
Li$_2$Sr[(Li$_{1-x}$Co$_x$)N]$_2$: $x = 0.3$\,\cite{Hoehn2021},
Li$_2$Sr[(Li$_{1-x}$Ni$_x$)N]$_2$: $x = 0.1$\,\cite{Cordier1989b},
Li$_2$Sr[(Li$_{1-x}$Cu$_x$)N]$_2$: $x = 0.39$\,\cite{Jager1992}).  
It is also interesting to note that the angle $\angle$N-$T$-N in the chain [(Li$_{1-x}T_x$)N] increases ($x = 0$: 174.5\,$^\circ$)\,\cite{Cordier1989b} with increasing $x$ for all $T$ to values between $175.5-177.5$\,$^\circ$. 
In contrast, Li$_2$Ca[(Li$_{1-x}$Mn$_x$)N]$_2$\,\cite{Klatyk2000} exhibits a different structure type.
\\

\section{Electrical transport}
\begin{figure}
\includegraphics[width=0.44\textwidth]{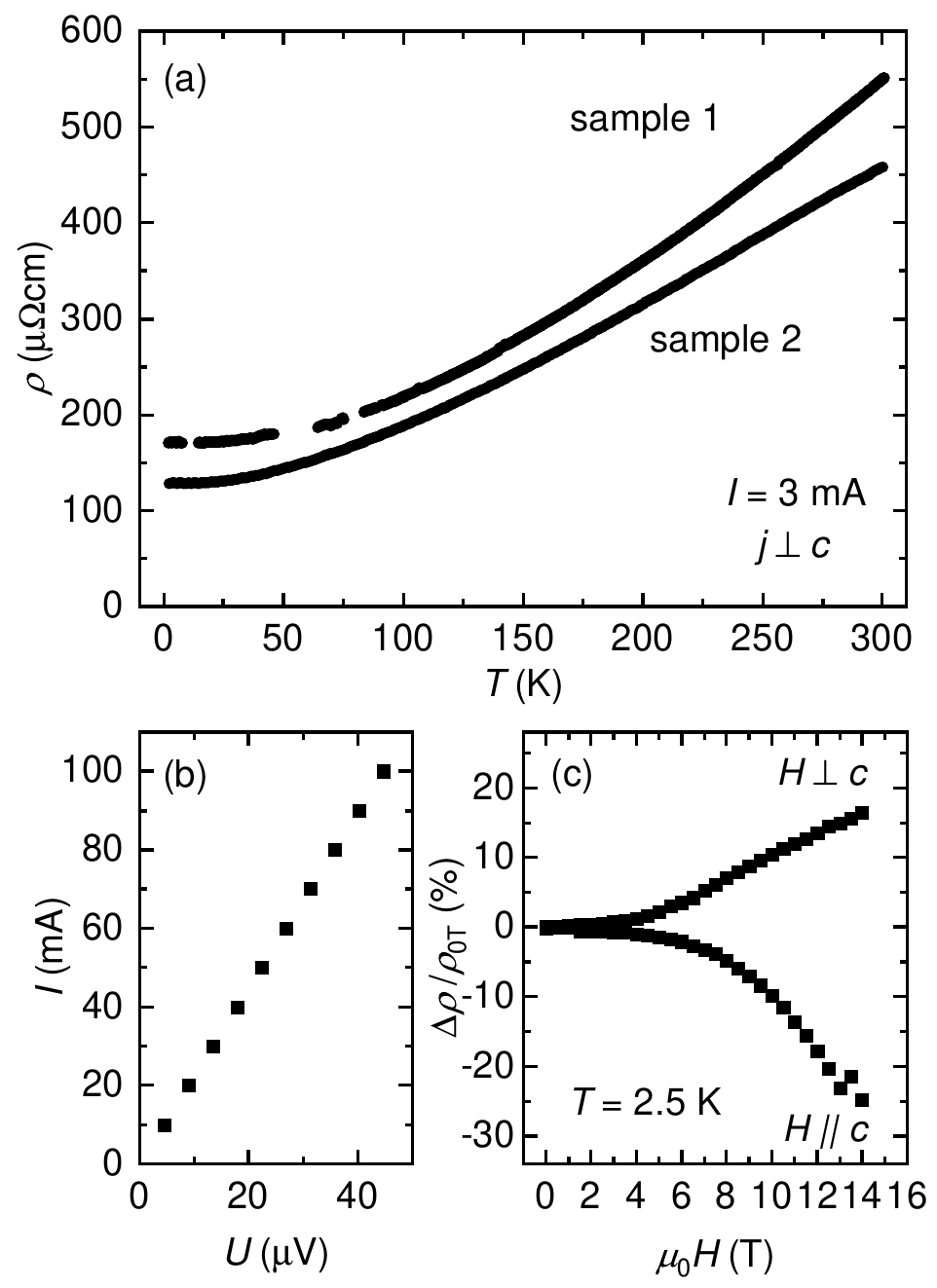}
\caption{\label{res} Electrical transport properties of \Mn\,obtained by using spring loaded, four wire contacts. (a) Linear temperature-dependence and relatively low absolute values at room temperature as key properties of metallic systems. (b) Ohmic current-voltage characteristics. (c) Pronounced anisotropy of the magneto-resistance at low temperatures. 
}
\end{figure}

Initial two-probe measurements using a standard multimeter indicated metallic behavior and motivated more accurate resistivity measurements as a function of temperature, excitation current, and magnetic field. 
Figure\,\ref{res}(a) shows the temperature-dependent electrical resistivity, $\rho(T)$ for $j \perp c$ measured on two different samples.
Absolute values amount to $\rho \sim 500\,\mu\Omega$cm at room temperature, a value that is larger than the one of elemental metals (for example $140\,\mu\Omega$cm for Mn).
Resistivities of this order of magnitude are somewhat smaller or comparable to what is referred to as 'bad metal'\,\cite{Emery1995}.
A linear temperature-dependence - a key property of a metallic material - is observed above $T \approx 200$\,K.
The residual resistivity ratio of $RRR > 3$ indicates an acceptable sample quality. 
The gaps in $\rho(T)$ at $T < 100$\,K that appear for sample 1 are caused by a temporary loss of contacts caused by thermal expansion and/or a mechanical blocking of the springs. 
Storing the measurement setup in a desiccator in between the runs helped to prevent those contact failures.

Ohmic current-voltage characteristics are found up to currents of at least 0.1\,A [Fig.\,\ref{res}(b)].
Using larger currents increases the risk of loosing the contacts presumably due to heating at the contact tips, which causes a local disintegration of the sample material. 

The magneto-resistance was measured for two different orientations of the crystal: $H \perp c$ with $j \parallel H$ and $H \parallel c$ with $j \perp H$. 
Figure\,\ref{res}(c) shows pronounced anisotropy and moderate absolute values obtained at $T = 2.5$\,K: the resistivity increases for field applied perpendicular to the $c$-axis but decreases for field applied parallel to $c$ (that is perpendicular to the $-$N$-$Mn$-$N$-$Mn$-$ chains).

\section{Magnetization}
\begin{figure}
\includegraphics[width=0.47\textwidth]{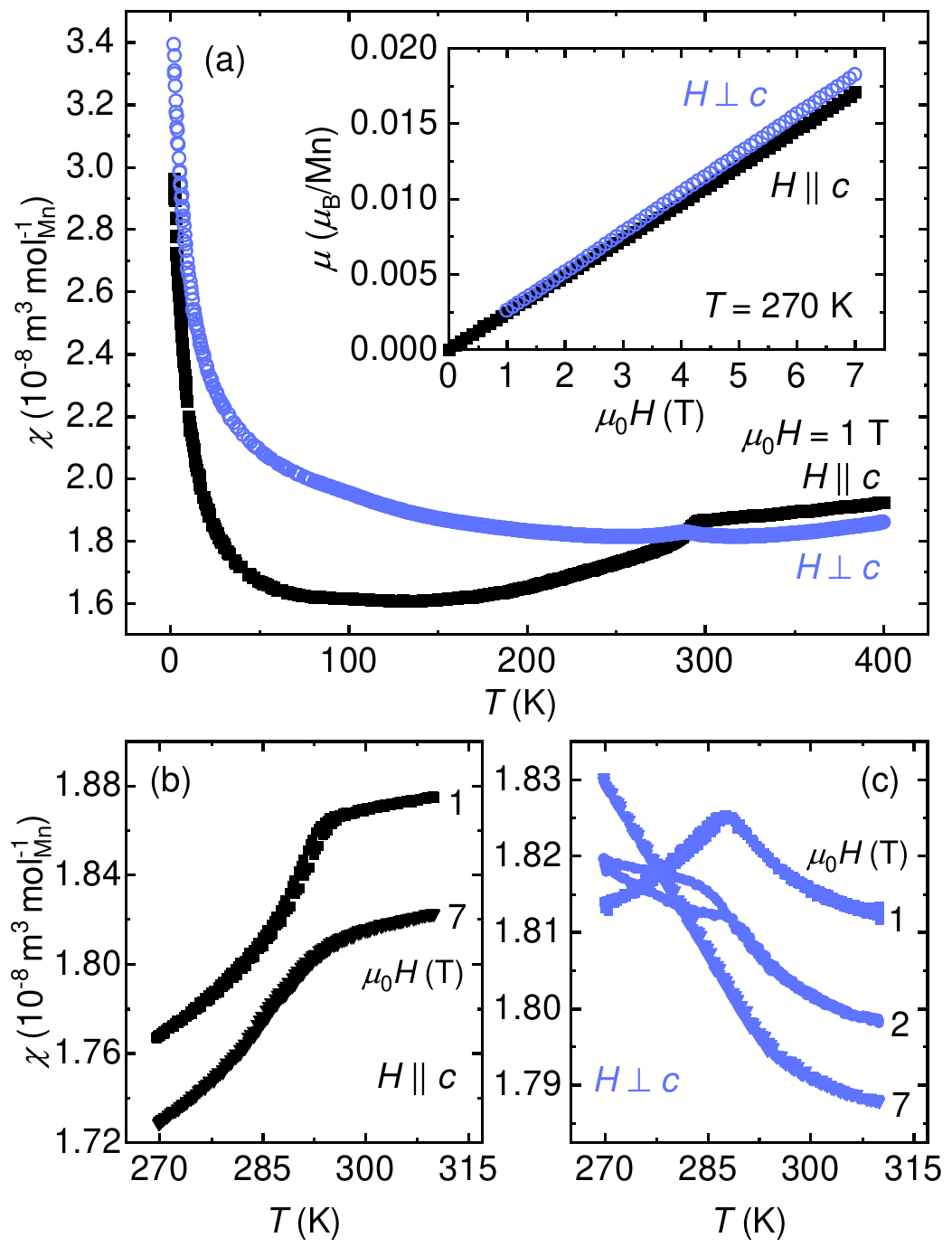}
\caption{\label{chi} Temperature- and field dependent magnetization of \Mn ($T > 2$\,K). (a) Magnetic susceptibility $\chi = M/H$ exhibits low anisotropy and temperature-dependence. The magnetization is linear in field (inset).
A clear anomaly is observed at $T_N = 290$\,K.
(b) $H \parallel c$: The step-like decrease in $\chi(T)$ at $T_N$ indicates antiferromangetic ordering in accordance with the anomaly shifting to lower temperature in larger applied fields.  
(c) $H \perp c$: a well-defined maximum is observed for $\mu_0H \leq 1$\,T at $T_N$. For larger applied fields $\chi(T)$ shows a monotonic increase upon cooling with a change of slope at $T_N$. 
}
\end{figure}

The magnetization was measured for two orientations: the applied field was aligned parallel ($H \parallel c$) or perpendicular ($H \perp c$) to the crystallographic $c$-axis. 
The obtained magnetic susceptibility $\chi(T) = M/H$ is plotted in Fig.\,\ref{chi}a. 
Nearly isotropic behavior is found for $T > 300$\,K with large absolute values and a linear increase of $\chi(T)$ with $T$.
The magnetization is linear in field for temperatures $T > 2$\,K as shown in the inset.
A significant increase of $\chi(T)$ for cooling below $T \sim 50$\,K indicates the presence of local magnetic moments (Curie tail).
 
A clear anomaly is observed around room temperature with the strongest change of slope at $T_N = 290$\,K ($H \parallel c$).
Antiferromagnetic (AFM) ordering is inferred from the step-like decrease upon cooling and the slight shift of the anomaly towards lower temperature for increasing applied fields [Fig.\,\ref{chi}(b), $H \parallel c$]. 
For $H \perp c$, a local maximum in $\chi(T)$ at $T = 288$\,K is found for $\mu_0H \leq 1$\,T.
As shown in Fig.\,\ref{chi}(c), a strong change of this behavior is observable in larger applied fields: a thermal hysteresis emerges and the anomaly gets broader for $\mu_0H = 2$\,T. 
A monotonous increase upon cooling is found for 3\,T$\leq \mu_0H \leq 7$\,T with the strongest change of slope close to $T_N$.

\section{Specific heat and magnetization at $T < 2$\,K}

\begin{figure}
 \includegraphics[width=0.48\textwidth]{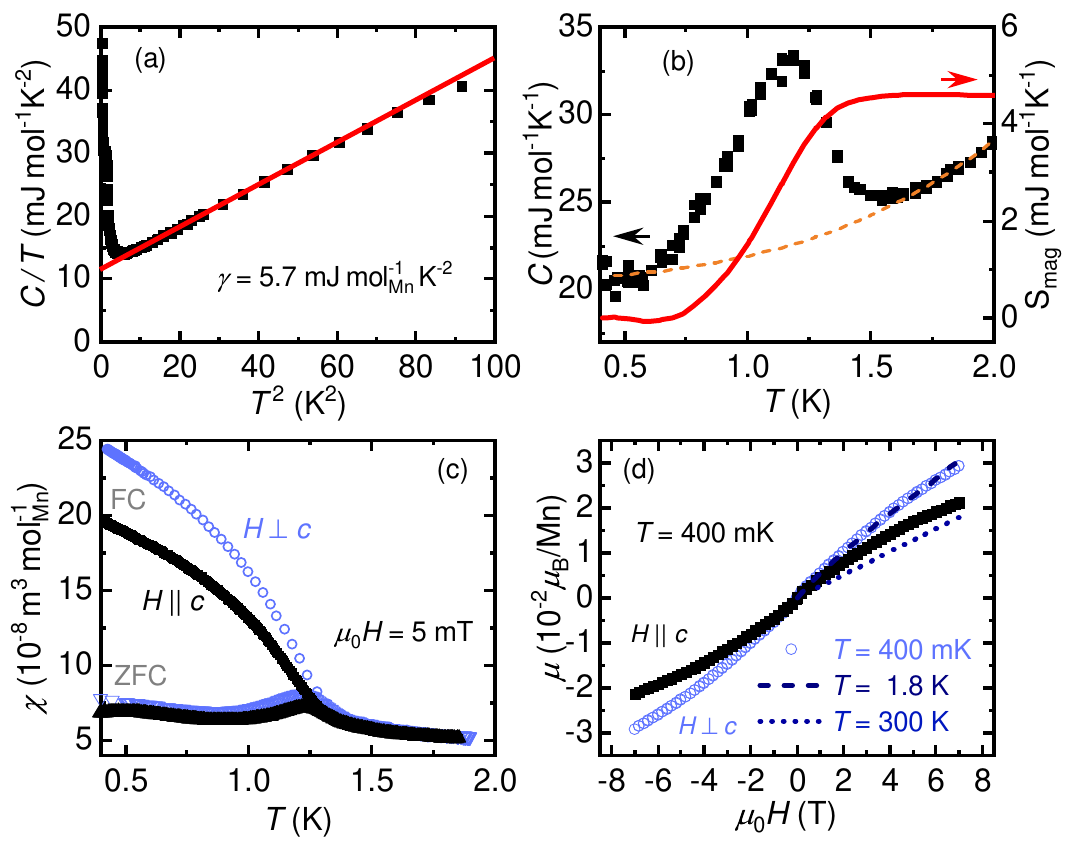}
\caption{\label{lowT} Specific heat $C(T)$ and magnetization of \Mn~for $T < 2$\,K. 
(a) Electron and phonon contribution lead to the expected linear behavior of $C/T$ as a function of $T^2$. The Sommerfeld coefficient is estimated to $\gamma = 5.7$\,mJ\,mol$^{-1}_{\rm Mn}$\,K$^{-2}$.  
(b) A broad maximum appears at $T_f = 1.2$\,K. 
The magnetic contribution was estimated by fitting a polynomial to the data points above and below the anomaly (dashed, orange line).  
The transition is associated with a significant amount of magnetic entropy $S_{\rm mag}$.
(c) The magnetic susceptibility $\chi(T) = M/H$ shows a maximum at $T_f$ in zero-field-cooled (ZFC) measurements for $H \perp c$ and $H \parallel c$. 
The field-cooled (FC) curves split from the ZFC at $T_f$ for both orientations.  
(d) Even at the lowest accessible temperature of $T = 400$\,mK, the field-dependent, isothermal magnetization is only slightly curved.
}
\end{figure}

Figure\,\ref{lowT}(a) shows the specific heat $C(T)$ plotted as $C(T)/T$ as a function of $T^2$. 
A linear dependence is observed for 4\,K $<T<10$\,K with a slightly enhanced Sommerfeld coefficient of $\gamma = 5.7$\,mJ\,mol$^{-1}_{\rm Mn}$\,K$^{-2}$.  
Towards lower temperatures, $C(T)$ increases and a comparatively broad peak is found at $T_f = 1.2$\,K [Fig.\,\ref{lowT}(b)].
In order to estimate the corresponding magnetic entropy, $S_{\rm mag}$, 
a 3rd order polynomial was fit to the data below $T = 0.6$\,K and above $T = 1.6$\,K (dashed line).
We find $S_{\rm mag} \approx 2.3$\,mJ\,mol$^{-1}_{\rm Mn}$\,K$^{-1}$.

Fig.\,\ref{lowT}(c) shows low temperature $\chi(T) = M/H$ for $H \parallel c$ and $H \perp c$. 
Similar to the whole investigated temperature range, only a small anisotropy is visible.
A maximum is found in zero-field-cooled (ZFC) measurements at $T_f = 1.2$ in very good agreement with $C(T)$. 
Field-cooled (FC) runs bifurcate from the ZFCs close to $T_f$.

Isothermal magnetization measurements at temperatures as low as $T = 400$\,mK are presented in Fig.\,\ref{lowT}(d). 
There is no substantial change for increasing the temperature above $T_f$ and further up to room temperature (dashed lines).
Even in the largest available fields, the magnetization amounts to less then 0.03\,$\mu_{\rm B}$ per Mn significantly below the saturation moment of $\mu_{\rm sat} = 4$\,$\mu_{\rm B}$ (for $S = 2$).
This proves the absence of (more than 1\,\%) localized Mn magnetic moments that are free or only weakly coupled for the whole temperature range investigated.

\section{Electronic Structure}
\begin{figure}
\includegraphics[width=0.48\textwidth]{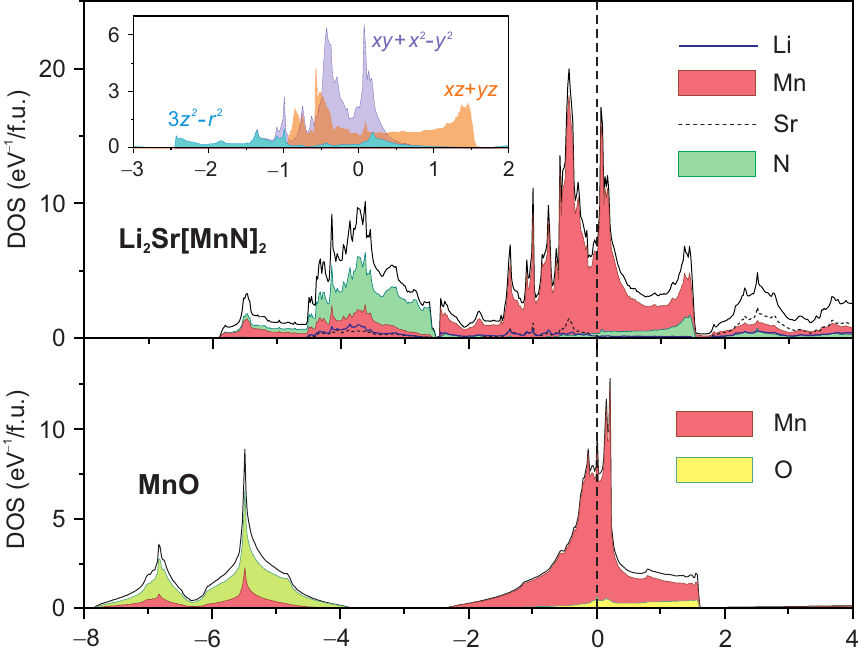}
\caption{\label{fig:dos}
LDA density of states for \Mn~(top panel) and MnO (bottom panel). The inset shows orbital-resolved DOS for Mn atoms in the former compound. The Fermi level is at zero energy.
}
\end{figure}

LDA density of states (DOS) for \Mn~features N $2p$ band between $-6$ and $-2.5$\,eV followed by the Mn $3d$ band that straddles the Fermi level (Fig.~\ref{fig:dos}). 
Only a tiny gap is observed between these bands, in contrast to low-valent Mn oxides, where Mn and O bands are usually well separated from each other. 
We choose MnO as the reference compound and compare its LDA DOS with that of \Mn. 
The shift of the $2p$ band toward higher energies is immediately visible in Fig.~\ref{fig:dos}. 
It can be quantified by the charge-transfer energy $\Delta$ that decreases from 4.1\,eV in the oxide to 2.4\,eV in the nitride. 
Here, $\Delta$ is estimated as the difference between centers of gravity of the Mn $3d$ and O/N $2p$ states.

Orbital-resolved DOS for \Mn~(Fig.~\ref{fig:dos}, inset) suggests the splitting of Mn $d$ states into crystal-field levels with $\eps_{3z^2-r^2}\simeq -0.98$\,eV, $\eps_{xy}\simeq\eps_{x^2-y^2}\simeq -0.32$\,eV, and $\eps_{yz}\simeq\eps_{xz}\simeq 0.30$\,eV, where the $z$ axis is chosen along the Mn--N bonds. 
This level scheme is consistent with the linear coordination of Mn$^{1+}$ and also shows that the charge-transfer energy for the $3z^2-r^2$ orbital is reduced to 1.3\,eV. 
This orbital then features a notably broader dispersion because of a much stronger hybridization with the N $2p$ states.

To probe the effect of electronic correlations on the band structure, we performed DFT+$U$ calculations for both MnO and \Mn. 
Experimental antiferromagnetic structure was chosen for the former, while in the latter case an antiferromagnetic structure with the highest symmetry was considered. 
It keeps the periodicity of the crystal structure and comprises ferromagnetic layers in the $ab$ plane coupled antiferromagnetically along the $c$ axis ($I\,4_1$ symmetry). 
Alternatively, a spin configuration with antiferromagnetic order along the Mn--N chains was constructed in a four-fold supercell and showed a very similar evolution of the band structure with $U$, so we focused on a simpler antiferromagnetic structure that allowed faster and more easily converged calculations at different values of $U$.

\begin{figure*}
\includegraphics[width=0.95\textwidth]{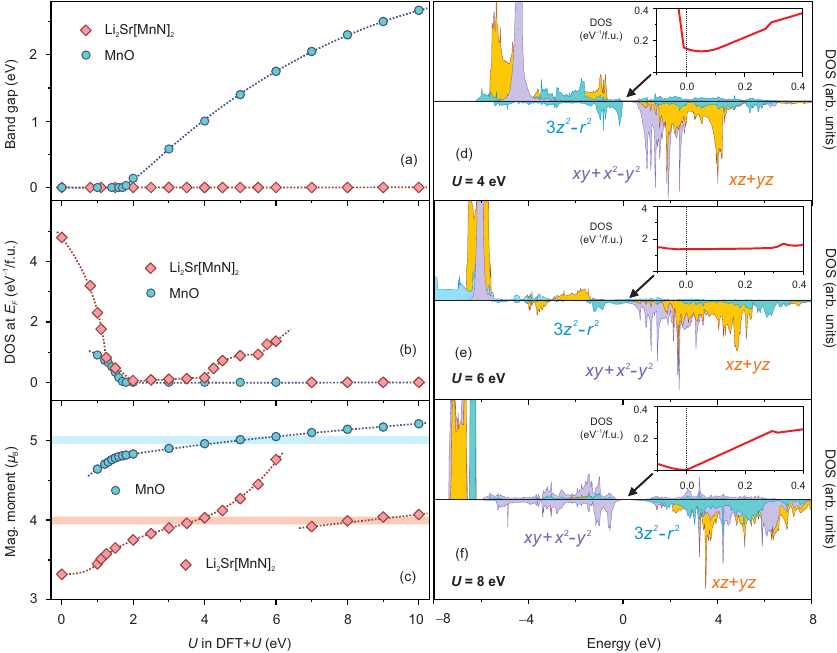}
\caption{\label{fig:u}
(a-c) Evolution of the electronic structure for MnO (circles) and \Mn~(diamonds) upon increasing the Coulomb repulsion parameter $U$ of DFT+$U$: band gap (a), total DOS at the Fermi level (b), and magnetic moment on Mn (c). (d-f) Orbital-resolved DOS for a single Mn atom in \Mn~calculated at $U=4$\,eV (d), $U=6$\,eV (e), and $U=8$\,eV (f). The Fermi level is at zero energy. The insets show total DOS in the vicinity of the Fermi level. The corresponding states are mostly of N $2p$ origin, but their exact position and, consequently, DOS at $E_F$, strong depend on the spin polarization of Mn $d$ orbitals.
}
\end{figure*}

MnO behaves as a typical Mott insulator (Fig.~\ref{fig:u}). A band gap opens at $U=1.8$\,eV and grows linearly, albeit with the gradually reduced slope at higher $U$, whereas magnetic moment on Mn systematically increases due to increased localization of the Mn $3d$ states. At high $U$ values, it even exceeds the nominal value of 5\,$\mu_B$ because of the enhanced polarization of the surrounding oxygen atoms. 

The evolution of \Mn~reveals several similarities. Its DOS at the Fermi level, $N(E_F)$, also decreases rapidly with $U$, whereas magnetic moment increases and changes slope around $U=1.8$\,eV, where $N(E_F)$ drops nearly to zero. The nominal magnetic moment of 4\,$\mu_B$ is reached at $U=4$\,eV, but in contrast to MnO, the gap never opens because a small fraction of N $2p$ states always remains at the Fermi level. This residual metallicity is due to the fact that the N $2p$ states are strongly mixed with Mn $3d$, especially the lowest-lying $d_{3z^2-r^2}$ orbital. It is occupied by the minority-spin electron of $3d^6$ Mn$^{1+}$ and remains unpolarized (Fig.~\ref{fig:u}d), so it can't be localized and gapped by adding $U$. Only at $U>4$\,eV does one observe the tendency toward the gradual localization of $d_{3z^2-r^2}$ states through their spin polarization. 

The steep increase in the magnetic moment of Mn above 4\,$\mu_B$ at $U>4$\,eV reflects this tendency toward polarizing the $d_{3z^2-r^2}$ states, with $N(E_F)$ increasing as the minority-spin states associated with $d_{3z^2-r^2}$ are shifted toward positive energies across the Fermi level (Fig.~\ref{fig:u}e). This process is completed at $U>6$\,eV, where the $d_{3z^2-r^2}$ states are fully polarized at the cost of placing minority-spin electron onto the $d_{x^2-y^2}$ orbital. 
The resulting configuration features a pseudogap with $N(E_F)=0$ but no gap opened even at very high $U=8-10$\,eV (Fig.~\ref{fig:u}f).

From the experimental $\gamma\simeq 5$\,mJ/mol\,K one expects $N(E_F)\simeq 2.1$\,eV$^{-1}$/f.u., which may be compatible with the weakly correlated scenario at $U\simeq 1$\,eV or with the intermediate regime at $U=5-6$\,eV (Fig.~\ref{fig:u}b), especially if renormalization of the effective mass and slight doping caused by the residual site mixing between Li and Mn are considered. 
Spectroscopic experiments and/or direct measurements of the local magnetic moment would be interesting to distinguish between these two possible regimes.

A more general outcome of our \textit{ab initio} study is the peculiar behavior of Mn$^{1+}$ nitride as a correlated material. 
Here, four $d$-orbitals show weaker hybridization with nitrogen and thus Mott behavior reminiscent of MnO, whereas the fifth $d$-orbital has a strong charge-transfer character that causes residual metallicity due to N $2p$ states remaining at the Fermi level even in the presence of strong correlations. 
Another unusual effect is the shift of the minority-spin electron from the lower-lying $d_{3z^2-r^2}$ orbital with the stronger charge-transfer character to the higher-lying $d_{x^2-y^2}$ orbital with the weaker charge-transfer character, because the latter orbital is easier to localize. 
This unconventional scenario reminds of some earlier findings for a Ni$^{1+}$ oxide in the strongly correlated limit~\cite{Lee2004} and may be a general mechanism of inducing metallicity at intermediate values of $U$. 
Overall, we find that Mn nitrides have a much stronger proclivity to the metallic behavior than undoped Mn oxides, and one $d$-orbital with the strong charge-transfer character is already sufficient to trigger it.

\section{Discussion}
Absolute values and temperature-dependence of $\rho(T)$ as well as DFT calculations indicate the presence of free charge carriers.
There is, however, no signature in $\rho(T)$ at $T_N$. 
If the proposed AFM ordering of Mn would be caused by a spin-density-wave transition, as observed for example in the parent compounds of Fe-based superconductors\,\cite{Klauss2008}, a clear anomaly is expected in $\rho(T)$. 
However, very similar transport behavior with no signature in $\rho(T)$ at the AFM ordering is observed in underdoped cuprates\,\cite{Ando2001} 
and LaMnO$_3$\,\cite{Brion1999}.
This is interpreted as a sign for spatial separation of conducting and magnetically ordering entities.
Nevertheless, we are going to discuss and disproof the following two hypothetical scenarios: 

a) the anomaly in $\chi(T)$ is caused by AFM ordering of an impurity phase.  
Given the large absolute values of $\chi(T)$, which exceed the nearly itinerant ferromagnet Pd more than a factor of 2\,\cite{Gerhardt1981}, only a local moment bearing material seems possible.
Such a phase, however, is expected to show Curie-Weiss behavior in the paramagnetic region above $T_N$, which is in contrast to the linear increase of $\chi(T)$ with increasing temperature observed experimentally [Fig.\,\ref{chi}(a)].
Obtaining a linear $\chi(T)$ as result of a superposition of two non-linear contributions (impurity Curie-Weiss and main phase with high $\chi$) seems highly unlikely.
Furthermore, the anisotropy of $\chi(T)$ with an increase for $H \parallel c$ and a decrease for $H \perp c$ upon heating above $T_N$ seems incompatible with a paramagnetic local moment scenario [Fig.\,\ref{chi}(b,c)]. 

b) the metallic transport is caused by a conducting impurity phase. 
The low absolute values of $\rho$ below m$\Omega$\,cm require a significant cross section of a potential current path even when highly conductive materials are considered. 
Optical microscopy on pristine as well as on polished surfaces never revealed indications for the presence of such a phase. 
Furthermore, exposure to air leads to a rapid increase of the resistivity by several orders of magnitude. 
We infer that also near-surface regions of \Mn~make an important contribution to the electrical transport and do not only connect spring loaded contacts and other metallic current paths. 
A lower air-sensitivity is expected if already insulating \Mn~merely bridges contact leads and a hypothetical conducting impurity phase.
 
Next, we focus on the high-temperature region ($T > 300$\,K) of $\chi(T)$. 
The large absolute values of $\chi \approx 1.9\cdot 10^{-8}$m$^3$mol$^{-1}_{\rm Mn}$ are very unusual, in particular in combination with the linear increase with temperature. 
The latter is normally not observed for local-moment-bearing or ferromagnetically ordered materials.
All other materials we are aware of, however, show significantly smaller absolute values of $\chi$.
A similar linear increase of $\chi(T)$ with temperature is found in Fe-based superconductors, though with lower absolute values at around room temperature (RT):
F-doped LaFeAsO with $\chi_{\rm RT} \approx 4 \cdot 10^{-9}$\,m$^3$mol$^{-1}$\,\cite{Klingeler2010}, 
BaFe$_2$As$_2$ with $\chi_{\rm RT} \approx 6 \cdot 10^{-9}$\,m$^3$mol$_{\rm Fe}^{-1}$\,\cite{Wang2009b, Klingeler2010}, 
or 
SrFe$_2$As$_2$\,\cite{Yan2008} and CaFe$_2$As$_2$\,\cite{Ronning2008} both with $\chi_{\rm RT} \approx 9 \cdot 10^{-9}$\,m$^3$mol$_{\rm Fe}^{-1}$.
The peculiar temperature-dependence was described by assuming a coexistance of localized and itinerant moments\,\cite{Kou2009}, frustrated local-moment model\,\cite{Schmidt2010}, or spin-density-wave scenario with strong antiferromagnetic fluctations above $T_{\rm N}$\,\cite{Zhang2009}.
Also (underdoped) cuprates show similar temperature-dependencies of $\chi(T)$ with even lower absolute values ($T > T_{\rm N}$), for example La$_{2-x}$Sr$_x$CuO$_4$ with $\chi_{\rm} \sim 1 \cdot 10^{-9}$\,m$^3$mol$^{-1}$\,\cite{Nakano1994}.
Note that $\chi$ of elemental Cr also increases with temperature with a roughly five times smaller slope and one order of magnitude lower $\chi_{\rm RT} = 2.4 \cdot 10^{-9}$\,m$^3$mol$^{-1}$\,\cite{Mcguire1952} when compared to \Mn.

Observing metallic conductivity in a nitride is not unprecedented, though with significant differences to \Mn:
Antiperovskite nitrides of $T_3X$N type\,\cite{Ettmayer1967} (with $X$ = \{Ga, Sn, $T$,...\}) do show metallic conductivity and magnetic ordering (for example Mn$_3$ZnN\,\cite{Kim2003, Sun2013}). 
Those structures, however, are based on transition-metal rich hosts in strong contrast to insulating, primarily ionic Li$_4$SrN$_2$.
More nitrogen-rich compounds for which a pronounced metallic temperature-dependence of $\rho(T)$ was experimentally observed are 
(SN)$_x$\,\cite{Walatka1973}, 
CaNiN\,\cite{Chern1990}, 
SrNiN\,\cite{Yamamoto1995}, 
Sr$_2$NiN$_2$\,\cite{Kowach2000}, and
Ca$_2$N\,\cite{Lee2013}.
No indications for magnetic ordering were found in any of those materials. 
  
Finally, we discuss the low-temperature properties with anomalies in $\chi(T)$ and $C(T)$ at $T_f = 1.2$\,K.
Magnetic ordering of $S = 2$ spins ($3d^6$ spin only, according to Mn$^{1+}$), lead to a change in entropy by $S_{\rm mag} = {\rm R}\ln 5 = 13.4$\,J\,mol$^{-1}_{\rm Mn}$\,K$^{-1}$.  
The measured magnetic entropy of $S_{\rm mag} \approx 2.3$\,mJ\,mol$^{-1}_{\rm Mn}$\,K$^{-1}$ amounts to only a small fraction of that value (and this does not change considering $S = 1/2$ instead of $S = 2$).
Ordering of 0.02\,\% of Mn or a correspondingly small amount of impurities are sufficient to cause the maximum in $C(T)$ and to account for the measured $S_{\rm mag}$. 
Furthermore, the effective magnetic moment obtained from the low-temperature Curie tail amounts to $\mu_{\rm eff} = 0.29(2)\,\mu_{\rm B}$ per Mn (for both orientations, 2\,K $ < T < 50$\,K, see Fig.\,\ref{chi}a).
Assuming that a certain fraction of Mn, $x_{\rm loc}$, carries local magnetic moments of $\mu_{\rm eff} \approx 4.9\,\mu_{\rm B}$, according to $S = 2$, leads to $x_{\rm loc} = 0.36$\,\%. 
[Note that $x_{loc} = (0.29/4.9)^2$].
Again, a small fraction of Mn or other local moment bearing impurities is sufficient to explain the observed behavior. 
The perfect agreement with the fraction of isolated Mn atoms of 0.36\,\%, which follows from $x = 0.94$, however, seems to be accidental. 
Those Mn have two Li as next-nearest neighbor along the chain and could be decoupled from the ordered states in longer $-$N$-$Mn$-$N$-$Mn$-$ chains. 
Observing an $S_{\rm mag}$ that is one order of magnitude smaller than expected for $x_{\rm loc} = 0.36$ could be caused by the large uncertainty in determining the non-magnetic background of $C(T)$ or neglecting crystal electric field effects and exchange interactions. 
It could also indicate magnetic frustration or spin-glass behavior with freezing of local moments at $T_f$\,\cite{Mydosh1993b}. 
Further similarities to spin-glasses are given by the splitting of FC and ZFC curves of $\chi(T)$ and the comparatively broad feature in $C(T)$.  
Investigating the anomaly at $T_f$ for varying $x$ could serve as a very sensitive method to probe the Mn concentration and stoichiometry of \Mn, provided isolated Mn do indeed cause the anomaly. 
In particular, the question arises of whether or not the anomaly disappears for $x \cong 1.00$. 
The assumption that the vast majority of Mn participates in itinerant, magnetic ordering at $T_{\rm N} = 290$\,K and only a small fraction of isolated magnetic moments (Mn or impurities) orders at $T_f = 1.2$\,K is supported by the preponderance of the obtained experimental data.

\section{Summary}
\Mn~single crystals of several millimeter along a side were grown from Li-rich flux.
Clear indications for itinerant antiferromagnetic ordering at $T_{\rm N} = 290$\,K are found.
The magnetic susceptibility above $T_{\rm N}$ increases linearly with temperature and amounts to an exceptionally large value of $\chi \approx 1.9\cdot 10^{-8}$m$^3$mol$^{-1}_{\rm Mn}$. 
The magnetic ordering is not accompanied by a structural transition as shown by low temperature single crystal X-ray diffraction.
Our main results can be understood based on structurally well-ordered but finite, linear chains of $-$N$-$Mn$-$N$-$Mn$-$ that are essential for the observed metallic electrical transport and magnetic properties. 
A small fraction of well below 1\% of isolated, local magnetic-moment-bearing Mn or a correspondingly small amount of a magnetic impurity orders/freezes at $T_f = 1.2$\,K.
The experimentally observed metallic resistivity is corroborated by LDA+U calculations that identified a strong mixing of N $2p$ with Mn $3d_{3z^2-r^2}$ states as 
the cause of metallicity even in the presence of sizable electronic correlations.
\Mn~is a unique magnetically ordered, metallic nitride with a highly unusual magnetic susceptibility that is reminiscent of the normal state of (underdoped) cuprate and iron-based superconductors. 

\section{Acknowledgments}
The authors thank Sascha Janster, Steffen Hückmann, Klaus Wiedenmann, and Alexander Herrnberger for experimental assistance.
German Hammerl is acknowledged for help with the electrical transport measurements.
This work was supported by the Deutsche Forschungsgemeinschaft (DFG, German Research Foundation) - grant number JE748/1.


%

\end{document}